# The Whole Warps the Parts: Attentional Certainty Modulates Ensemble-Induced Bias in Visualizations

Jennifer E. Corbett and Brian W. Anthony

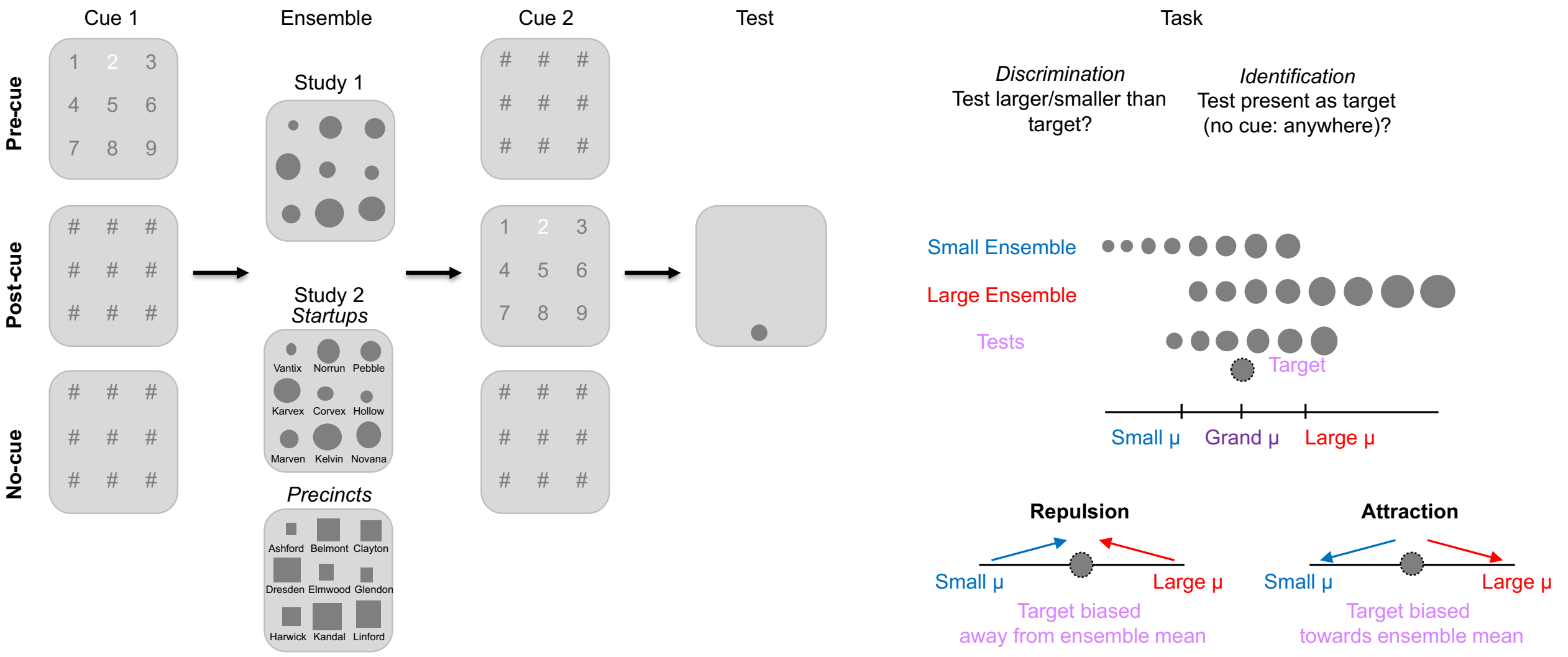

**Fig. 1. The same value can appear more extreme or more typical depending on what viewers are asked to remember.** We measured how surrounding data bias judgments of a physically identical target embedded in small- (blue) or large-mean (red) ensembles. A target was specified before (pre-cue) or after (post-cue) the ensemble, or never specified (no-cue). Study 1 used circle arrays for identification and discrimination; Study 2 extended identification to startup valuations and precinct vote contributions. Target-specific judgments favored repulsion away from the ensemble mean, whereas uncued judgments produced attraction toward it.

***Abstract*— Data visualizations often require viewers to extract both ensemble statistics and individual values from collections of graphical marks. Yet individual items embedded within an ensemble can be systematically biased by that context. Prior studies have reported both repulsion away from and attraction toward ensemble statistics, but these effects have been observed across different conditions. We test whether task or attentional certainty explains these opposing forms of ensemble-induced bias. In Study 1, observers briefly viewed ensembles of differently sized circles and then judged the size of an individual circle embedded within the ensemble. Using a common stimulus-generation framework, we tested both identification and discrimination when the target was specified before or after the ensemble, and an additional no-cue identification condition without a specified target. Under the shared cued conditions, effects were predominantly repulsive or near zero in both identification and discrimination. In contrast, eliminating target-specific information reversed ensemble-induced bias from repulsion to attraction. Identification precision also decreased systematically from pre-cue to post-cue to no-cue conditions. Study 2 replicated this cue-dependent pattern in graphical displays of startup valuations and precinct vote contributions. Our findings isolate attentional certainty as a key determinant of whether ensemble context repels or attracts individual-item representations, extending the design question beyond precision to how attention-directing choices determine which bias the viewer will be operating under.**

## I. INTRODUCTION

Visualizations routinely ask viewers to do two things at once: extract the overall structure of a dataset and recover specific values from within it. Ensemble perception supports the first operation by allowing observers to rapidly represent statistical properties of sets[1–3], a capacity also used to extract aggregate properties from graphical displays[4,5]. But ensemble representations can also distort their individual constituents, producing either attraction toward ensemble statistics[6–8] or repulsion away from them[9,10]. What determines the direction of this ensemble-induced bias remains unresolved. Here, we test two candidate mediators with direct relevance to visualization: task and attentional certainty.

As illustrated in Fig. 1, observers judged individual values embedded among other marks, first in abstract arrays of differently sized circles and then in graphical displays representing startup valuations and precinct vote contributions. We used a common psychophysical framework that allowed us to manipulate the ensemble surrounding a physically identical target and estimate its perceived value from responses across multiple test sizes. Within this general framework, we manipulated both task and attentional certainty. The task-dependent reversal between identification and discrimination we predicted was not observed. Instead, bias direction changed when target-specific information was removed, such that cued identification judgments favored repulsion from the surrounding ensemble, whereas uncued judgments produced attraction toward it. Identification precision also decreased systematically with attentional certainty. The same patterns generalized to graphical displays. Together, our results show that attention-directing choices can affect not only how precisely an individual value is read, but also the direction in which surrounding data bias that reading.

## II. BACKGROUND

### *A. Ensemble Encoding and Unequal Weighting*

Ensemble statistics can be extracted rapidly even when individual items are poorly represented[2]. These summaries span low- and high-level visual features, space, time, multiple summary statistics, memory, and neural mechanisms[3]. Even before the term “ensemble encoding” was used, Healey and colleagues[11] introduced the idea that knowledge of the viewer’s ability to rapidly extract summary information about entire sets of items could offer important advantages when visualizing information. Later studies explicitly proposed ensemble encoding as the mechanism underlying the rapid extraction of complex information such as the strength and magnitude of correlations in scatterplots[12,13] and the average position of clusters of points in multiclass scatterplots[4].

However, ensemble representations do not necessarily weight every constituent equally. Outlier discount occurs when values that differ substantially from the rest of a set contribute less to its perceived summary[14–16]. Perceptual organization also influences which elements contribute to a summary[17]. Conversely, amplification occurs when attended, salient, or otherwise prioritized elements receive greater weight[18–21]. Related weighting biases occur in visualization. Scatterplot means are pulled toward larger or darker marks[22], line-graph averages toward regions of greater variability[23], and outlying scatterplot points can be underweighted in estimates of central tendency and dispersion[24]. Overall, we have a comparatively coherent account of how individual values contribute to ensemble representations. The complementary influence remains an outstanding question: How does the ensemble alter the perceived values of individual constituents?

### *B. Ensemble-Induced Bias on Individual Values*

Ensemble statistics can produce repulsive effects, shifting individual-item judgments away from the enveloping ensemble statistic[9,10,25–27]. Likewise, visualization research demonstrates that contextual properties can determine the direction of systematic error. For example, McColeman et al.[28] found that while adding reference context to bars and dots could improve reproduction precision, this simultaneously introduced repulsion away from an implicit categorical boundary.

Ensemble context can instead produce attractive biases for individual items, shifting reported values toward the mean of the surrounding set[6,7,29,30]. Observers are also more likely to endorse previously unseen items near an ensemble mean or prototype and more likely to reject items outside the ensemble range[2,8,31–35]. Attractive contextual effects have also been reported for graphical marks, with remembered target locations displaced toward surrounding marks or perceptual groups [36].

Fig. 2 illustrates these opposing effects using representative results from the prior literature. Together, these findings establish that contextual biases can be attractive or repulsive, and that properties of a display or judgment can determine which pattern emerges. The unresolved question is which property governs that direction.

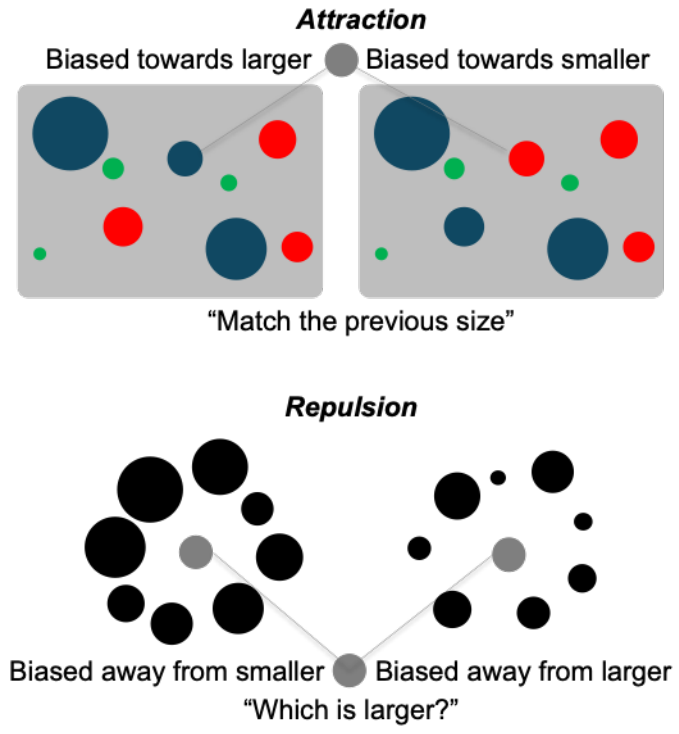

**Fig. 2. Representative examples of repulsive and attractive ensemble-induced bias.** Bottom: Repulsion: Simultaneous size-contrast example based on the Ebbinghaus illusion, in which identically-sized central items appear smaller/larger as an inverse function of the larger/smaller mean size of the surrounding context. Top: Attraction: The remembered size of an individual item is biased toward the statistical properties of the same color large/small ensemble[6].

### *C. Task Type as a Candidate Mediator*

One potential explanation for the opposing directions of ensemble-induced bias comes from the tasks used to measure individual-item perception. Several prominent demonstrations of repulsion have used tasks requiring discrimination of a test value following ensemble adaptation or relative to an individual target in a surrounding context. These include ensemble-adaptation paradigms in which observers discriminate the magnitude of a subsequently presented item following adaptation to an ensemble (e.g., size[10]; duration[25]; attractiveness[37], as well as when a test item must be discriminated relative to a target embedded among surrounding reference items[9]. Such tasks may favor differentiation of the target from its surrounding context.

By contrast, attractive effects have been reported most often in tasks requiring identification or continuity of an individual representation over time. These include adjusting a test item to reproduce a previously viewed item[6,7,29,30] and identifying whether or which item appeared in an earlier ensemble[2,8,32,35]. This potential dissociation points to the observer’s task a plausible mediator of bias direction. However, repulsive and attractive effects have generally been studied with different stimuli, procedures, and attentional conditions, making it difficult to determine whether task itself caused the opposing biases.

This possibility is directly relevant to visualization because graphical displays support tasks that place different demands on item-specific versus ensemble representations. Szafir et al.[5] identify four broad operations supported by ensemble coding in visualization tasks varying systematically in how strongly they require individuation versus integration: identification, summarization, segmentation, and structure estimation. Identification emphasizes access to particular values, whereas summarization explicitly depends on integration across the display; segmentation and structure estimation involve intermediate forms of organization such as grouping and trend extraction. If ensemble-induced repulsion and attraction depend on similar task demands, the direction of perceptual bias could change with what the viewer is asked to do with the same underlying display.

We test this possibility directly by holding the stimulus-generating framework constant (including the target value, ensemble distributions, timing, and test values) while varying whether observers perform identification or discrimination. Both tasks are tested under the same pre- and post-cue conditions. Therefore, Study 1 provides a controlled test of task as a candidate determinant of ensemble-induced bias with direct relevance to the different operations viewers perform on visualized data.

### *D. Attentional Certainty as a Candidate Mediator*

A second possible explanation is attentional certainty about the item being evaluated. Classic cueing studies show that advance cues allow attention to be allocated to relevant information during encoding, whereas post-cues allow relevant information to be selected only after initial encoding[38–40]. In ensemble perception, this distinction maps onto a broader tradeoff between focused attention to an individual object and distributed attention across a set. For example, Chong and Treisman[41] found better size averaging when attention was distributed broadly across a display than when a concurrent search task required serially focused processing of individual elements.

Cueing also changes how selectively information within an ensemble can be represented. Chong and Treisman[42] initially reported little cost when the relevant color-defined subset was specified only after presentation, but later work showed that their displays could partly be judged from the whole-set mean. With stimuli requiring genuine subset selection, pre-cueing improved subset judgments, whereas whole-set representations remained comparatively robust[43–45]. At the individual-item level, Utochkin and Brady[30] found attractive biases toward the ensemble mean in individual-item memory, whereas Choi and Golomb[9] found repulsion when targets were distinguished by color and when they were identified by a post-cue. Notably, they tested targets smaller or larger than all surrounding reference items, whereas the present study tests a physically constant, intermediate-sized item embedded among both smaller and larger values. Taken together, these findings demonstrate that ensemble-induced attraction and repulsion can occur despite target-specific information and that cueing changes the balance between item-specific and ensemble information, but do not establish a simple relationship between target certainty and the direction of ensemble-induced bias.

Visualization research provides a parallel reason to expect. attentional certainty to matter. Viewers selectively filter marks before forming aggregate judgments in multiclass scatterplots[4] miss conspicuous patterns contained within task-irrelevant subsets of a visualization[46], and are guided toward particular comparisons by spatial grouping, color, and other attention-directing cues[47]. Explicit highlighting and annotation similarly increase memory for emphasized patterns in explanatory visualizations[48]. These findings establish that attentional guidance changes what viewers select, compare, and retain, but not whether it also determines how the surrounding ensemble biases the selected value itself.

Towards these ends, we manipulate attentional certainty directly within the same stimulus-generating framework. Study 1a provides an initial manipulation by identifying the target either before the ensemble (pre-cue) or after it (post-cue), in both identification and discrimination. Study 1b repeats these conditions in both tasks and extends the certainty manipulation by adding a no-cue identification condition, in which no individual target is ever specified and observers instead judge whether the test item had appeared anywhere in the ensemble. By extending the manipulation from differences in the timing of target-specific information to its complete absence, this provides a stronger test of whether attentional certainty determines both the precision and direction of ensemble-induced bias.

### *E. Contributions*

Our work makes four main contributions.

**An empirical resolution of a fundamental question in human visual perception.** Whether ensemble context repels or attracts individual-item representations has remained unresolved, in part because the two effects have typically been observed across studies that also differed in stimuli, procedures, tasks, and attentional demands. Our design removes much of that ambiguity by using a common stimulus-generating framework while experimentally varying only the candidate mediators themselves. This controlled comparison showed that changing task did not produce the predicted reversal in bias direction, whereas manipulating certainty did. Together, our findings reconcile previously conflicting results and constrain mechanistic accounts of how ensemble and individual-item representations interact.

**An applied demonstration in graphical displays of practical interest.** We show that the same qualitative pattern extends from abstract stimuli to graphical displays representing startup valuations and precinct vote contributions. This establishes a direct link between a longstanding problem in ensemble perception and the perceptual interpretation of individual data values.

**A reframing of the visualization design problem.** Our results suggest that attention-directing design choices should not be considered solely in terms of whether they improve selection, precision, or efficiency. They can also alter the direction of systematic bias in the perceived value itself. In fact, we did not identify a cueing regime that consistently eliminated ensemble influence. Therefore, visualization interactions may be understood not simply as tools for reducing error, but as choices that can change the form that error takes. The relevant design question becomes not only how accurately a value is read, but under which type of contextual bias that value is read.

**An interdisciplinary contribution between vision science and visualization research.** Our approach advances a broader effort to bring psychophysical theories and methods into visualization research[49], while using visualization-relevant stimuli to address an open problem in basic perception. Ensemble-perception theory provides hypotheses about how graphical marks are represented, while visualization provides applied contexts in which those theories can be tested and refined. We see this reciprocal use of basic and applied perceptual research as an important opportunity for both fields.

## III. STUDY 1: ABSTRACT DISPLAYS

### *A. Study 1 Overview*

Study 1 comprised two data collections using a common within-subjects paradigm. As illustrated in Fig. 1, each trial consisted of a first cue, a nine-item ensemble, a second cue, and a test circle. Observers either identified whether the test had appeared at a specified target location (identification) or discriminated whether it was larger or smaller than the target (discrimination). Attentional certainty was manipulated by specifying the target location before (pre-cue) or after (post-cue) the ensemble, or by providing no target-specific information and asking whether the test had appeared anywhere in the set (no-cue). Study 1a crossed both tasks with pre- and post-cueing, while Study 1b additionally included no-cue identification. No-cue discrimination was excluded because, without a specified target, larger/smaller judgments would concern the ensemble distribution or range rather than the perceived size of an individual target[33].

Ensemble mean size and test size provided a common framework for measuring contextual bias. Each nine-item display contained a physically identical 1.05° target surrounded by eight circles with either a small (0.85°) or large (1.25°) mean size. Test size varied around the target (seven levels in Study 1 and five in Study 2), allowing us to estimate the response functions for each ensemble condition. For identification, we fit Gaussian functions to the proportion of "Present" responses, with the peak indexing the test size most likely to be endorsed as appearing at the specified location or, under no-cueing, anywhere in the ensemble. For discrimination, we fit logistic functions to the proportion of "Larger" responses, with the point of subjective equality (PSE) estimating the test size perceived as equal to the target.

As illustrated in Fig. 3, differences between the small- and large-mean functions index the direction and magnitude of ensemble-induced bias. Repulsion shifts judgments away from the ensemble mean, yielding negative large-minus-small differences in peak or PSE. Attraction reverses this pattern, yielding positive differences. The fitted functions also provide independent measures of precision, such that smaller Gaussian width (sigma) and steeper logistic slopes indicate greater precision. Together, these measures allow us to test whether task or attentional certainty determines the direction of ensemble-induced bias and whether precision decreases as certainty decreases.

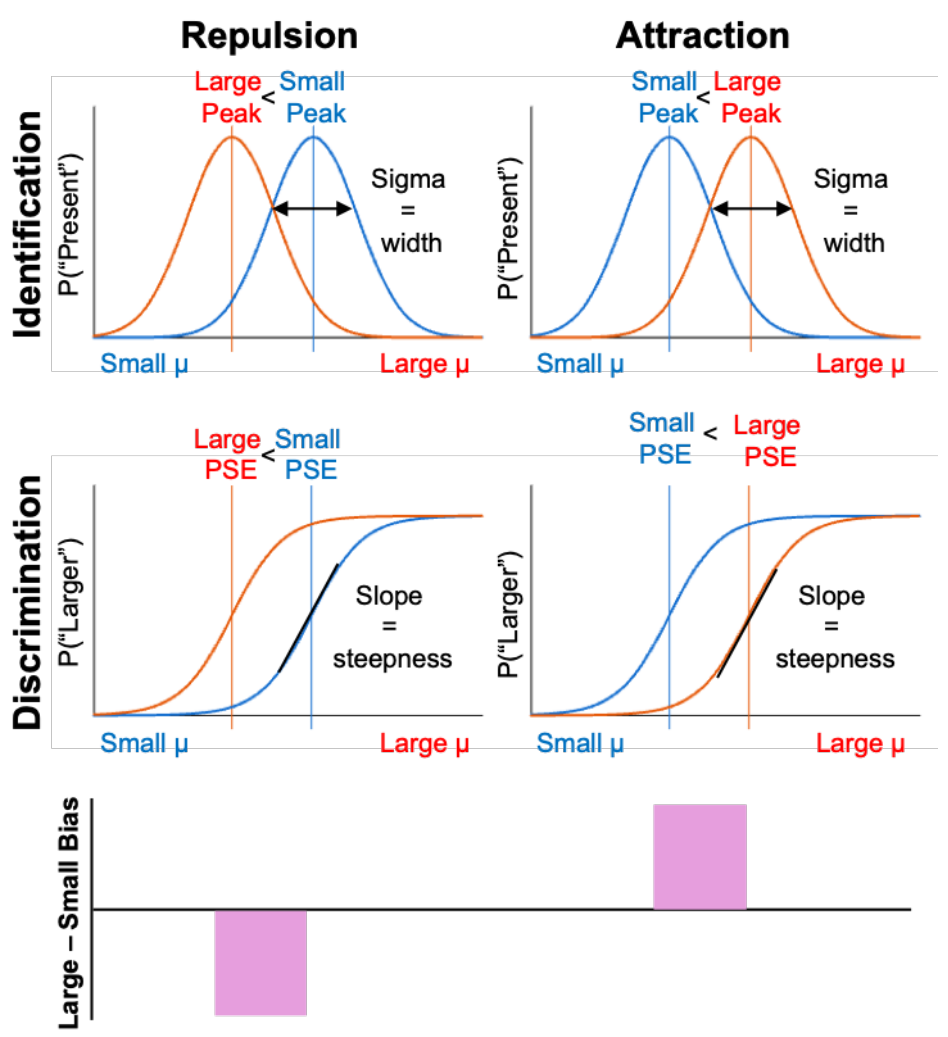

**Fig. 3. Psychometric signatures of ensemble-induced repulsion and attraction.** Identification (top) responses are modeled with Gaussian functions relating test size to the proportion of "Present" responses. The fitted peak estimates bias (perceived target size) and the width, sigma, indexes precision. Discrimination (middle) responses are modeled with logistic functions relating test size to the proportion of "Larger" responses. The inflection point, PSE, estimates bias and the slope indexes precision. The physical target is identical in the small- and large-mean ensemble conditions. For ensemble-induced repulsion (left), the target is perceived as larger in the small-mean ensemble and smaller in the large-mean ensemble, yielding negative large-minus-small peak/PSE values (bottom). For ensemble-induced attraction (right), the ordering reverses, resulting in positive large-minus-small mean values.

We tested three central hypotheses.

**H1: Task determines the direction of ensemble-induced bias.** If task determines bias direction as suggested by results from the ensemble literature, then identification should produce attractive bias with a smaller peak for small- versus large-mean ensembles indexed by positive large-minus-small peak differences illustrated in Fig. 3, and discrimination should produce repulsive bias with a larger PSE for small- versus large-mean ensembles indexed by negative large-minus-small PSE differences.

**H2: Attentional certainty determines the direction of ensemble-induced bias.** If attentional certainty determines bias direction, then decreasing certainty should shift from the repulsive toward the attractive bias illustrated in Fig. 3 with the largest magnitude of large-minus-small repulsion in the pre-cue condition, weaker or intermediate effects in the post-cue condition, and attraction in the no-cue identification condition.

**H3: Greater attentional certainty produces a more precise individual-item representation.** If precision increases with attentional certainty, then identification Gaussian fits should widen as in Fig. 3 and sigma should increase from pre-cue to post-cue to no-cue conditions, whereas discrimination logistic fits should have steeper slopes under pre-cueing than post-cueing.

*B. Study 1 Methods*

*Participants.* Study 1a: 24 participants. Study 1b: 23 participants, none of whom had participated in Study 1a. An additional participant in Study 1b was excluded because the person discontinued the session before completing the entire experiment All participants reported normal or corrected-to-normal vision and gave informed consent under a protocol approved by MIT COUHES institutional review board. In Study 1a, participants were between 20 and 33 years old (M = 25.8), 18 were female, and 6 were male. In Study 1b, participants were between 18 and 32 (M = 23.4), 15 were female, and 8 were male. Participants were recruited via the SONA online recruitment system from MIT's undergraduate and graduate student population and the greater Boston undergraduate community and received $10 in Amazon vouchers upon completion.

*Apparatus.* The experiment ran in MATLAB with Psychtoolbox 3[50,51]. Participants viewed the 59.90 x 34.00 cm computerized display at a resolution of 2560 x 1440 pixels from a distance of approximately 57 cm, such that one degree of visual angle corresponded to about 43 pixels.

*Stimuli.* Each ensemble display contained a nine-circle ensemble arranged in a 3 x 3 grid centered on the screen (grid cells 2° of visual angle on a side, ensemble region ~6° square). Circles were dark gray on a mid-gray background. Each circle's position was jittered by ±0.1° on every trial to prevent low-level retinal or early cortical adaptation effects across the many trials of each block[52]. On each trial, the target location was chosen randomly from one of the nine possible locations and contained a circle that was physically identical across small-mean and large-mean displays as the grand mean (1.05°). The remaining eight positions were filled with circles pseudorandomly selected without replacement from the eight possible sizes in the small-mean set (0.5, 0.6, 0.7, 0.8, 0.9, 1.0, 1.1, 1.2°; mean 0.85°) or the large-mean set (0.9, 1.0, 1.1, 1.2, 1.3, 1.4, 1.5, 1.6°; mean 1.25°). As illustrated in Fig. 1, the two size sets are symmetric around grand mean target size and share four sizes (0.9, 1.0, 1.1, 1.2°), equating substantial low-level display properties across the two ensemble conditions. After the ensemble and a brief interstimulus interval, a single test circle appeared at a ±0.1°jittered position 4° below the ensemble. Across both mean sizes, test sizes were 0.8, 0.9, 1.0, 1.05, 1.1, 1.2, and 1.3°.

*Trial sequence.* Each trial began with a 0.5° fixation cross that remained in the center of the screen until the participant pressed the spacebar. The first cue appeared for 500 ms, followed by a 250-ms blank interval, the ensemble for 500 ms[53,54], another 250-ms blank, and the second cue for 500 ms. Informative cues displayed numbered positions (1–9), with the target location in white and the others in dark gray. Uninformative cues displayed dark gray "#" symbols at all positions. The informative cue appeared first in pre-cue trials, second in post-cue trials, and not at all in no-cue trials. Finally, the test circle appeared below the ensemble until response. Participants used the up/down arrow keys to indicate whether the test was present/absent at the specified target location (cued identification), present/absent anywhere in the set (no-cue identification), or larger/smaller than the target (discrimination).

*Design.* Task, cueing condition, ensemble mean, and test size were varied within subjects. Study 1a comprised four task-by-cue blocks and Study 1b five, with block order counterbalanced across participants. Each block was preceded by written and illustrated instructions and task-specific practice trials. Within each block, two ensemble means and seven test sizes were crossed with 10 repetitions per cell (140 trials), plus five catch trials in which the target was one of the ensemble set items rather than the grand mean. Catch trials were included to help ensure that participants did not become aware of the otherwise constant target and were excluded from analysis. Study 1a therefore comprised 580 trials and Study 1b 725 trials per participant. Sessions lasted approximately 45–60 minutes.

*C. Study 1 Analysis*

We fit separate psychometric functions for each participant and combination of task, cueing condition, and ensemble mean size. Identification responses were fit with a three-parameter Gaussian:

$$P(\text{Present} \mid x) = A \exp[-(x - \mu)^2 / (2\sigma^2)]$$

where A is the amplitude, μ the peak, and σ the width. Discrimination responses were fit with a four-parameter logistic:

$$P(\text{Larger} \mid x) = L + (U - L) / [1 + \exp(-k(x - \mu))]$$

where L and U are the lower and upper asymptotes, μ the PSE, and k the slope. Asymptotes were bounded to prevent unrealistic lapse rates. Fits were obtained by chi-squared minimization using five starting parameter sets, retaining the solution with the smallest chi-squared value.

We analyzed the location (peak/PSE) and precision (sigma/slope) parameters separately to evaluate the corresponding hypotheses. Because peak and PSE derive from different psychometric functions, we did not combine identification and discrimination in a single ANOVA. Instead, H1 was evaluated from the predicted directions of ensemble-induced bias within each task under the shared pre- and post-cue conditions.

*Bias:* Raw peak or PSE values were entered into repeated-measures ANOVAs with Attentional Certainty and Ensemble Mean as within-subjects factors, separately for each task. Study 1a and merged Study 1 used 2 (Certainty: pre-cue, post-cue) × 2 (Ensemble Mean: small, large) ANOVAs. Study 1b used a 3 × 2 ANOVA for identification and a 2 × 2 ANOVA for discrimination. The Ensemble Mean main effect tested overall contextual bias, whereas the Certainty × Ensemble Mean interaction tested whether that bias changed with attentional certainty (H2). We had no corresponding hypothesis for the Certainty main effect. Planned one-sample t-tests evaluated the within-participant large-minus-small peak or PSE differences against zero for each cue condition, and paired-samples t-tests compared these differences across cue conditions.

*Precision:* Because H3 concerned attentional certainty rather than differences between ensemble means, sigma and slope were averaged across the small- and large-mean conditions within each participant and cue condition. These averages were analyzed using one-way repeated-measures ANOVAs on Attentional Certainty, with two levels for Study 1a, merged Study 1, and Study 1b discrimination, and three levels for Study 1b identification. Planned paired-samples t-tests compared precision across cue conditions.

We analyzed raw peak and PSE values rather than deviations from the physical target (1.05°), because general (individual and mean) size overestimation could otherwise be conflated with ensemble-induced bias[18,20,55–58]. Moreover, the small- and large-mean ensembles shared four item sizes, with no categorical boundary between them. Therefore, the theoretically relevant contrast was the within-participant difference between identical targets embedded in the two ensemble contexts.

*D. Study 1 Results*

Fig. 4 summarizes the fitted psychometric functions and corresponding measures of ensemble-induced bias and precision across the Study 1 conditions.

*Study 1a (Fig. 4; Left panel) The predicted task-dependent reversal in bias direction was not observed*

*Bias:* For identification, there was an overall effect of Ensemble Mean on peak, $F(1,23) = 8.12$, $p = .009$, $\eta_p^2 = .261$, indicating significant ensemble-induced bias across the two cueing conditions. The Certainty × Ensemble Mean interaction was not reliable, $F(1,23) = 1.69$, $p = .207$, $\eta_p^2 = .068$. Planned tests of the Large-minus-Small peak difference showed repulsion in both conditions. Pre-cueing produced a significant negative bias, M = -0.026°, SE = 0.004, $t(23) = -6.71$, $p < .001$, dz = -1.37. Post-cueing also produced a negative bias, M = -0.071°, SE = 0.034, $t(23) = -2.07$, $p = .050$, dz = -0.42. The magnitude of the bias did not differ reliably between pre- and post-cue, $t(23) = 1.30$, $p = .207$, dz = 0.27.

Discrimination did not show the predicted opposite pattern. The overall effect of Ensemble Mean on PSE was not reliable, $F(1,23) = 0.72$, $p = .406$, $\eta_p^2 = .030$, and there was no Certainty × Ensemble Mean interaction, $F(1,23) = 1.81$, $p = .192$, $\eta_p^2 = .073$. Nevertheless, pre-cueing showed a small negative PSE difference that approached significance, M = -0.016°, SE = 0.008, $t(23) = -2.07$, $p = .050$, dz = -0.42, whereas post-cueing showed essentially no bias, M = 0.002°, SE = 0.013, $t(23) = 0.13$, $p = .895$, dz = 0.03. The two cueing conditions did not differ reliably, $t(23) = -1.34$, $p = .192$, dz = -0.27.

Taken together, the predicted task-dependent reversal from identification attraction to discrimination repulsion was not observed. Instead, all reliable ensemble-induced bias effects in Study 1a were repulsive.

*Precision*: Identification precision depended strongly on attentional certainty. Sigma was substantially smaller for the pre-cue, M = 0.103, SE = 0.005, than post-cue condition, M = 0.252, SE = 0.028, $F(1,23) = 33.85$, $p < .001$, $\eta_p^2 = .595$. The planned comparison confirmed this difference, $t(23) = -5.82$, $p < .001$, dz = -1.19, indicating substantially greater precision when the target was specified before the ensemble.

Discrimination precision showed no corresponding effect. Mean slope was 59.30 for pre-cue and 53.35 for post-cue, and the difference was not reliable, $F(1,23) = 0.43$, $p = .518$, $\eta_p^2 = .018$, $t(23) = 0.66$, $p = .518$, dz = 0.13.

*Study 1b (Fig. 4; Middle panel): Removing target certainty reversed the direction of ensemble-induced bias*

*Bias:* Study 1b replicated the pre- and post-cue identification pattern from Study 1a and extended the manipulation by adding the no-cue condition. Critically, ensemble-induced bias changed substantially across the three certainty conditions, as reflected in a strong Certainty × Ensemble Mean interaction, $F(2,44) = 21.00$, $p < .001$, $\eta_p^2 = .488$. The main effect of Ensemble Mean was not reliable, $F(1,22) = 1.02$, $p = .323$, $\eta_p^2 = .044$, because the opposing repulsive and attractive effects across cue conditions largely cancelled when averaged together.

Planned comparisons revealed the theoretically important change in direction. Pre-cueing again produced repulsion, M = -0.026°, SE = 0.006, $t(22) = -4.08$, $p < .001$, dz = -0.85. Post-cueing was also negative in direction but did not differ reliably from zero, M = -0.057°, SE = 0.031, $t(22) = -1.82$, $p = .083$, dz = -0.38. In contrast, no-cueing produced significant attraction, M = 0.140°, SE = 0.033, $t(22) = 4.28$, $p < .001$, dz = 0.89.

The two cued conditions did not differ from one another, $t(22) = 1.10$, $p = .283$, dz = 0.23. In contrast, the no-cue condition differed strongly from both pre-cue, $t(22) = -5.24$, $p < .001$, dz = -1.09, and post-cue conditions, $t(22) = -5.23$, $p < .001$, dz = -1.09. Therefore, reducing target certainty from pre- to post-cue conditions did not itself reverse bias direction. The reversal only emerged in the no-cue condition when target-specific information was removed entirely.

The discrimination conditions again included only pre- and post-cue, and broadly replicated Study 1a. Pre-cueing produced reliable repulsion, M = -0.027°, SE = 0.010, $t(22) = -2.85$, $p = .009$, dz = -0.59, whereas post-cueing remained near zero, M = 0.006°, SE = 0.017, $t(22) = 0.36$, $p = .720$, dz = 0.08. The Certainty × Ensemble Mean interaction was not reliable, $F(1,22) = 2.57$, $p = .123$, $\eta_p^2 = .105$, and the pre- and post-cue bias scores did not differ reliably, $t(22) = -1.60$, $p = .123$, dz = -0.33.

*Precision*: Identification precision also varied systematically with attentional certainty, $F(2,44) = 20.77$, $p < .001$, $\eta_p^2 = .486$. Sigma increased monotonically as certainty decreased, from pre-cue, M = 0.103, SE = 0.006, to post-cue, M = 0.253, SE = 0.036, to no-cue conditions, M = 0.349, SE = 0.035. Pre-cueing was more precise than post-cueing, $t(22) = -4.53$, $p < .001$, dz = -0.94, and more precise than no-cueing, $t(22) = -7.17$, $p < .001$, dz = -1.50. The post-cue versus no-cue comparison was in the same predicted direction and approached significance, $t(22) = -2.05$, $p = .052$, dz = -0.43.

Discrimination precision again showed no reliable certainty effect. Mean slope was 52.48 under pre-cueing and 60.82 under post-cueing, $F(1,22) = 0.46$, $p = .503$, $\eta_p^2 = .021$, $t(22) = -0.68$, $p = .503$, dz = -0.14.

Together, the Study 1b results reveal two related effects of attentional certainty in identification. As certainty decreased, estimates became progressively less precise, and when target-specific information was absent entirely, ensemble-induced bias reversed from repulsion to attraction.

*Merged Study 1: Robustness of the shared pre- and post-cue conditions*

Because Studies 1a and 1b used independent samples but shared the same pre- and post-cue conditions, we pooled these overlapping conditions to obtain more precise estimates across 47 participants.

*Bias*: For identification, the merged sample showed a strong overall effect of Ensemble Mean on peak, $F(1,46) = 13.71$, $p < .001$, $\eta_p^2 = .230$. The Certainty × Ensemble Mean interaction was not reliable, $F(1,46) = 2.95$, $p = .093$, $\eta_p^2 = .060$. Both cueing conditions showed reliable repulsion. Pre-cueing produced a mean Large-minus-Small peak difference of -0.026°, SE = 0.004, $t(46) = -7.07$, $p < .001$, dz = -1.03. Post-cueing also produced significant repulsion, M = -0.064°, SE = 0.023, $t(46) = -2.79$, $p = .008$, dz = -0.41. However, their difference did not reach significance, $t(46) = 1.72$, $p = .093$, dz = 0.25.

For discrimination, the merged sample clarified the pre-/post-cue pattern. Pre-cueing produced reliable repulsion, M = -0.021°, SE = 0.006, $t(46) = -3.51$, $p = .001$, dz = -0.51, whereas post-cueing remained near zero, M = 0.004°, SE = 0.010, $t(46) = 0.37$, $p = .713$, dz = 0.05. The Certainty × Ensemble Mean interaction was reliable in the larger sample, $F(1,46) = 4.38$, $p = .042$, $\eta_p^2 = .087$, and the pre- versus post-cue bias comparison was correspondingly reliable, $t(46) = -2.09$, $p = .042$, dz = -0.31.

Therefore, the merged analysis reinforced the main conclusion that identification and discrimination did not exhibit the predicted task-dependent reversal. Reliable ensemble-induced effects were predominantly repulsive, with discrimination shifting from pre-cue repulsion toward approximately zero under post-cue.

*Precision*: The identification precision effect was especially robust in the merged sample. Mean sigma increased from 0.103 under pre-cueing to 0.252 under post-cueing, $F(1,46) = 52.55$, $p < .001$, $\eta_p^2 = .533$. The corresponding planned comparison was significant, $t(46) = -7.25$, $p < .001$, dz = -1.06.

Discrimination precision did not vary with cueing. Mean Slope was 55.96 for pre-cue and 57.01 for post-cue, $F(1,46) = 0.02$, $p = .891$, $\eta_p^2 < .001$, $t(46) = -0.14$, $p = .891$, dz = -0.02.

*E. Study 1 Summary*

The Study 1 results provide distinct outcomes for our three hypotheses:

**H1: If task determines the direction of ensemble-induced bias, then identification should produce attraction whereas discrimination should produce repulsion.** This hypothesis was not supported. Across Study 1a, Study 1b, and the merged analysis of the shared pre- and post-cue conditions, changing from identification to discrimination did not produce the predicted reversal in bias direction. Reliable effects in the cued conditions were generally repulsive or near zero in both tasks.

**H2: If attentional certainty determines the direction and magnitude of ensemble-induced bias, then decreasing certainty should shift bias from repulsion toward attraction.** This hypothesis was supported most clearly in Study 1b. Pre-cue identification produced significant repulsion, post-cue remained on the repulsive side but did not differ reliably from zero, and no-cue produced robust attraction. The no-cue bias differed significantly from both cued conditions. Therefore, decreasing certainty from pre- to post-cueing alone was not sufficient to reverse bias direction, whereas removing all target-specific information produced the predicted crossover from repulsion to attraction.

**H3: If greater attentional certainty produces more precise individual-item representations, then identification sigma should increase as certainty decreases and discrimination slope should be steeper under pre-cueing than post-cueing.** This hypothesis was strongly supported for identification. Sigma was consistently smaller under pre-cue than post-cue, and Study 1b showed the predicted graded increase from pre-cue to post-cue to no-cue. In contrast, discrimination slope did not differ reliably between pre- and post-cue conditions.

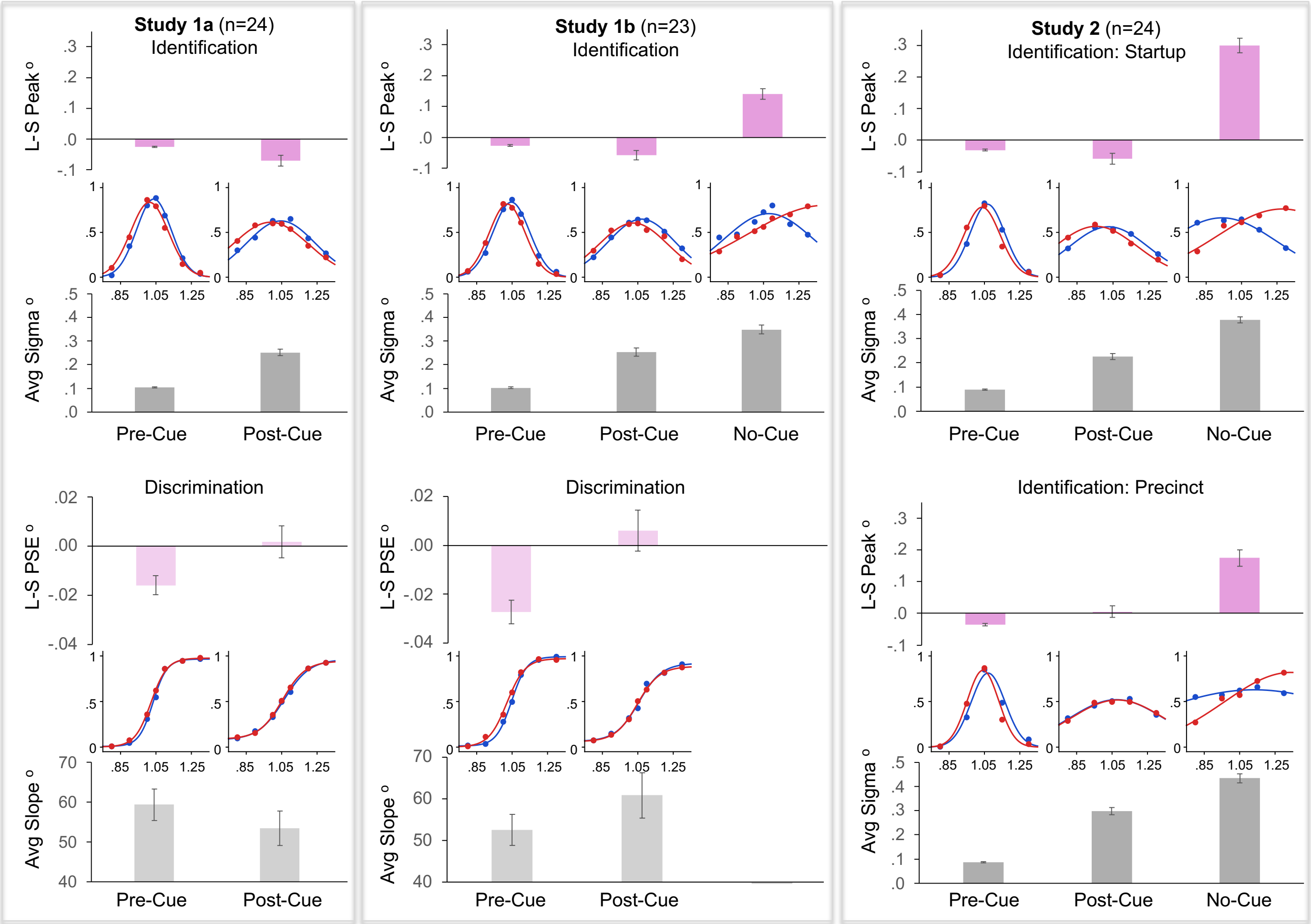

**Fig. 4 Ensemble-induced bias and precision across Study 1a (left), Study 1b (middle), and Study 2 (right).** Study 1 includes identification (top) and discrimination (bottom). Study 2 includes startup valuation (top) and precinct vote-count identification (bottom). Violet bars show large-minus-small ensemble differences in fitted peak (identification) or PSE (discrimination), with negative values indicating repulsion and positive values attraction. Insets show fitted psychometric functions for small-mean (blue) and large-mean (red) ensembles. Gray bars show identification width (sigma, lower indicates greater precision) or discrimination slope (higher indicates greater precision). Error bars represent the conditional SEMs. Across identification conditions, pre-cueing generally produced repulsion, no-cueing produced attraction, and precision decreased from pre-cue to post-cue to no-cue.

## IV. STUDY 2: BUBBLE AND BAR CHARTS

### *A. Study 2 Overview*

Study 2 tested whether the certainty-dependent identification pattern observed with abstract circles in Study 1 generalizes to graphical displays with real-world semantic framing. We retained only the identification task because Study 1 showed that the full range of attentional certainty required to produce a reversal in bias direction could be tested only with identification. Specifically, the no-cue condition provided the low-certainty endpoint where repulsion reversed to attraction.

Although the ensemble perception literature gives no particular reason to expect ensemble-induced bias to differ between circles and rectangular bars[3], graphical perception depends on the visual channel used to encode magnitude. Circle size is primarily conveyed through area, whereas bar values are conveyed through position and length, channels that differ in perceptual precision[59]. Bar judgments can also be systematically biased by incidental properties like aspect ratio[60], and contextual reference structure can introduce repulsion in reproduced graphical values[28]. Therefore, changing from circles to bars provides a stronger generalization test than simply changing the semantic framing of the same stimuli. Along these lines, we tested two graphical scenarios in Study 2 (Fig 1). The startup scenario preserved the same circular marks and size-encoding dimension used in Study 1, but framed the items as startup valuations and the judgment as verification of a reported value. The precinct scenario provided a further generalization test by framing the values as precinct vote counts and replacing circles with vertical rectangular bars whose heights encoded magnitude.

We tested two additional hypotheses in Study 2:

**H4: The certainty-dependent identification pattern generalizes to graphical displays.** If the Study 1 identification pattern generalizes to semantically framed graphical displays, then decreasing attentional certainty should shift ensemble-induced bias from repulsion toward attraction, with negative large-minus-small peak differences under pre-cue, a weaker or intermediate effect under post-cue, and positive differences under no-cue. Identification precision should also decrease with certainty, reflected by increasing sigma from pre-cue to post-cue to no-cue conditions.

**H5: The certainty-dependent pattern generalizes across mark geometry and encoding channel.** If these ensemble-induced effects are not specific to the circular stimuli used in Study 1, then the same qualitative pattern of bias and precision should emerge when magnitude is represented by rectangular bars in the precinct scenario.

*B. Study 2 Methods*

*Participants.* 24 participants completed Study 2. Six had previously participated in Study 1a and five had participated in Study 1b. Participants were between 18 and 33 years old (M = 23.9); 18 were female and 6 were male. Recruitment, compensation, and apparatus matched Study 1.

*Stimuli and procedure.* Study 2 followed the same trial sequence and cueing procedure as Study 1. In the startup scenario, nine circles were arranged in the same 3 × 3 configuration used in Study 1, with fictional startup names assigned to the items and circle size representing valuation. In the precinct scenario, nine vertical rectangular bars were arranged in the same spatial configuration, with fictional precinct names assigned to the items and bar height representing vote count. Bar width was held constant at 1.04° of visual angle so that magnitude varied only along the vertical dimension. To minimize low-level differences among the text labels, names were drawn from fixed-length fictional name pools within each scenario and randomly assigned to item positions on each trial.

The small- and large-mean ensembles and physical target value matched Study 1. Test values were reduced to five levels (0.80°, 0.95°, 1.05°, 1.15°, and 1.30°). Each trial began with a 500-ms cue display showing the nine item names at their corresponding display locations, followed by a 250-ms interval, the 500-ms ensemble display with each name shown below its corresponding mark, a second 250-ms interval, and a second 500-ms cue display. In the pre-cue condition, one name was highlighted in white on the first cue display, identifying which startup or precinct would later be queried and the second cue display was uninformative, with all names gray. In the post-cue condition, all names were gray on the first cue display, so participants viewed the ensemble without knowing which item would be queried and the relevant item name was then highlighted in white on the second cue display. In the no-cue condition, all names remained gray on both cue displays and no specific item was ever identified.

Finally, a single test mark appeared below the ensemble and remained until response together with a source-attributed verification question. For cued startup trials, the question took the form, "[source] reports this as [startup]'s valuation. Is that correct? (UP=Yes / DOWN=No)," and for precinct trials, "[source] reported that [precinct] contributed this many votes. Is that correct? (UP=Yes / DOWN=No)." In the no-cue condition, the item name was omitted so that the question concerned whether the displayed magnitude had occurred anywhere in the preceding set: "[source] reported a precinct contributed this many votes. Is that correct?" or "[source] reports this as a startup's valuation. Is that correct?".

*Design and analysis.* Scenario, cueing condition, ensemble mean, and test size were varied within subjects. Each participant completed six experimental blocks: the three cue-condition blocks for one scenario followed by the three cue-condition blocks for the other scenario. Participants completed all three pre-cue, post-cue, and no-cue blocks within a scenario before moving to the second scenario rather than alternating between scenarios. Scenario order was counterbalanced across participants, and the order of the three cue-condition blocks within each scenario was independently counterbalanced across the six possible cue orders.

Within each block, the two ensemble means and five test sizes were fully crossed with 10 repetitions per cell, yielding 100 experimental trials. Each block also contained four catch trials (excluded from analysis) in which the target was one of the ensemble set items rather than the grand-mean target. Also as in Study 1, each block was preceded by written and illustrated instructions and task-specific practice trials.

Study 2 used the same identification fitting and analysis procedures as Study 1b, with peak indexing perceived item size and sigma indexing precision. Each scenario was analyzed separately using a 3 (Certainty: pre-cue, post-cue, no-cue) × 2 (Ensemble Mean: small, large) repeated-measures ANOVA on peak. Planned one-sample tests again evaluated the large-minus-small peak difference against zero within each cue condition, and planned paired tests compared bias across cue conditions. Sigma was again averaged across small- and large-mean ensembles and analyzed with a one-way repeated-measures ANOVA across the three certainty conditions. H5 was evaluated by whether the predicted certainty-dependent pattern was also present in the precinct scenario rather than from a direct statistical comparison of null results between scenarios.

*C. Study 2 Results*

The right panel in Fig. 4 illustrates Study 2's results alongside Study 1. As in Study 1b identification, the critical measures are the large-minus-small peak difference indexing bias direction and sigma indexing precision.

*Startup: Semantic framing preserved the Study 1 pattern*

*Bias.* The startup scenario provided the most direct generalization test because the physical marks remained circles and only their semantic interpretation and response framing changed. Ensemble-induced bias varied strongly with attentional certainty, reflected in a large Certainty × Ensemble Mean interaction, $F(2,46) = 40.34$, $p < .001$, $\eta p^2 = .637$. There was also an overall effect of Ensemble Mean, $F(1,23) = 11.89$, $p = .002$, $\eta p^2 = .341$. The cue-specific effects closely reproduced Study 1b. Pre-cue produced significant repulsion, M = -0.031°, SE = 0.005, $t(23) = -5.81$, $p < .001$, dz = -1.19. Post-cue remained repulsive in direction but did not differ reliably from zero, M = -0.058°, SE = 0.034, $t(23) = -1.70$, $p = .103$, dz = -0.35. In contrast, no-cue produced strong attraction, M = 0.300°, SE = 0.045, $t(23) = 6.67$, $p < .001$, dz = 1.36. Pre- and post-cue did not differ reliably, $t(23) = 0.74$, $p = .466$, dz = 0.15. No-cue differed strongly from both pre-cue, $t(23) = -7.07$, $p < .001$, dz = -1.44, and post-cue, $t(23) = -7.32$, $p < .001$, dz = -1.49. Therefore, changing the abstract identification question to a startup-valuation judgment preserved the same certainty-dependent reversal observed in Study 1b.

*Precision.* Precision also showed the same graded relationship with certainty, $F(2,46) = 51.37$, $p < .001$, $\eta p^2 = .691$. Sigma increased from pre-cue, M = 0.089, SE = 0.006, to post-cue, M = 0.226, SE = 0.024, to no-cue, M = 0.377, SE = 0.027. All planned comparisons were significant: pre-cue versus post-cue, $t(23) = -6.06$, $p < .001$, dz = -1.24; pre-cue versus no-cue, $t(23) = -10.22$, $p < .001$, dz = -2.09; and post-cue versus no-cue, $t(23) = -4.51$, $p < .001$, dz = -0.92.

Together, the startup results support H4. The certainty-dependent reversal in bias and accompanying decrease in precision generalized from the abstract identification task to a semantically framed graphical judgment while preserving the same circular marks.

*Precinct: The pattern generalized across graphical encodings*

*Bias.* The precinct scenario tested whether the same pattern would persist when magnitude was represented by rectangular bars rather than circles. The certainty-dependent change in bias remained evident, with a significant Certainty × Ensemble Mean interaction, $F(2,46) = 8.84$, $p < .001$, $\eta p^2 = .278$, and an overall effect of Ensemble Mean, $F(1,23) = 5.63$, $p = .026$, $\eta p^2 = .197$. Pre-cue again produced reliable repulsion, M = -0.036°, SE = 0.006, $t(23) = -5.83$, $p < .001$, dz = -1.19. Post-cue was essentially unbiased, M = 0.005°, SE = 0.037, $t(23) = 0.14$, $p = .889$, dz = 0.03. No-cue again produced significant attraction, M = 0.174°, SE = 0.051, $t(23) = 3.41$, $p = .002$, dz = 0.70. Pre- and post-cue did not differ reliably, $t(23) = -1.04$, $p = .309$, dz = -0.21. No-cue differed from both pre-cue, $t(23) = -4.08$, $p < .001$, dz = -0.83, and post-cue, $t(23) = -2.61$, $p = .016$, dz = -0.53. Therefore, replacing circles with rectangular bars did not change the qualitative certainty-dependent pattern of ensemble-induced bias.

*Precision.* Precision again decreased systematically with attentional certainty, $F(2,46) = 48.60$, $p < .001$, $\eta p^2 = .679$. Sigma increased from pre-cue, M = 0.086, SE = 0.005, to post-cue, M = 0.299, SE = 0.031, to no-cue, M = 0.433, SE = 0.036. All three planned comparisons were significant: pre-cue versus post-cue, $t(23) = -7.25$, $p < .001$, dz = -1.48; pre-cue versus no-cue, $t(23) = -10.00$, $p < .001$, dz = -2.04; and post-cue versus no-cue, $t(23) = -3.25$, $p = .004$, dz = -0.66.

The Precinct results therefore support H5. The certainty-dependent shift from repulsion toward attraction and the graded loss of precision generalized across a change in mark geometry and graphical encoding.

### D. Study 2 Summary

The Study 2 results supported both generalization hypotheses.

**H4: If the certainty-dependent identification pattern generalizes to graphical displays, then decreasing certainty should shift ensemble-induced bias from repulsion toward attraction while reducing precision.** This hypothesis was supported in both scenarios. With the same circular marks used in Study 1 but a semantically framed startup-valuation judgment, pre-cueing produced repulsion, post-cueing remained repulsive but was not significantly different from zero, and the no-cue condition produced strong attraction. Sigma also increased systematically across pre-cue to post-cue to no-cue conditions.

**H5: If the certainty-dependent pattern generalizes across mark geometry and encoding channel, then the same qualitative pattern should emerge when circular area-based marks are replaced by rectangular bars.** This hypothesis was also supported across scenarios. In the precinct scenario, pre-cueing again produced repulsion, post-cueing was near zero, and the no-cue condition produced attraction. Precision showed the same graded pre-cue < post-cue < no-cue ordering. Overall, the results of Study 2 show that the Study 1 patterns of ensemble-induced bias on identification judgements generalized not only to semantically meaningful graphical judgments, but also across a change in the basic graphical encoding used to represent magnitude.

## V. Discussion

Across three data collections, we found that ensemble-induced bias depended primarily on how clearly an individual target was specified. When a target was identified before or after the ensemble, judgments generally showed repulsion or negligible bias, whereas removing target-specific information produced significant attraction toward the ensemble mean. Identification precision also decreased systematically as target certainty decreased, with sigma increasing from pre-cue to post-cue to no-cue conditions. Study 2 replicated both the reversal in bias direction and the graded loss of precision when circle-size judgments were framed as startup valuations and when circles were replaced by bars representing precinct vote counts. Overall, our findings demonstrate that attentional certainty modulates the direction and magnitude of ensemble-induced bias in abstract and real-world visualization contexts.

### A. Complementary Mechanisms of Ensemble-Item Interactions

Our findings suggest a complementary process to the manner in which selective attention alters individual item contributions to ensemble representations. Even without explicitly identifying outliers, observers discount extreme or deviant values when estimating ensemble statistics[14,15,61]. Conversely, attended, salient, or reward-associated items can receive greater weight in ensemble estimates[19–21]. Choi and Chong[18] further demonstrated that cueing an individual item shifted perceived mean size toward that item's size, with stronger effects under pre-cueing. Our findings suggest a complementary influence in the opposite ensemble-to-item direction. When no individual target was specified, set-membership judgments were attracted toward the ensemble mean, whereas judgments about specified targets generally showed ensemble-induced repulsion effects that were also strongest under pre-cueing. Together, these findings raise the possibility that attention regulates both directions of ensemble–item interactions. Outlier discounting may preserve a stable ensemble representation when individual items are not selected, with attraction toward the mean reflecting greater reliance on that representation. Conversely, attentional amplification of individual contributions to the ensemble, together with repulsion of specified targets from their surrounding context may help preserve the distinctiveness of behaviorally relevant items.

Utochkin et al.'s[62] population-response model provides a potential computational framework for such complementary ensemble-item attentional effects. Individual feature signals are pooled into a population response from which ensemble statistics are decoded. Greater activation near the center of the feature distribution naturally reduces the influence of outliers, supporting robust averaging without explicitly identifying or excluding individual elements. Utochkin et al.[62] propose that attention could modulate the weighting of individual signals or the tuning of pooling neurons, while feedback between ensemble- and item-level representations could support outlier detection. These mechanisms offer a possible basis for linking robust ensemble encoding with the selective amplification and differentiation of individual items. Extending it to account for how attentional selection changes the influence of pooled ensemble information on individual-item judgments could test whether these opposing biases arise from related mechanisms.

Our precision results are consistent with this interpretation. Across identification conditions, sigma was smallest when the target was specified before the ensemble, increased when it was specified afterward, and was largest when no target was specified. Although these differences cannot isolate noise arising during encoding, maintenance, retrieval, or decision-making, they indicate that reduced target-specific information coincided with poorer precision and a shift from repulsion toward attraction. This pattern is consistent with increasing reliance on the pooled ensemble representation as individual-item information becomes less accessible.

### B. Implications for Visualization

Information visualization increasingly exploits observers' ability to rapidly extract ensemble statistics from complex displays[5,49]. Yet accurately representing a distribution does not guarantee accurate judgments about its individual values. Previous research has demonstrated how graphical design choices can mislead interpretation[63], even when visualizations accurately represent the underlying data[64]. Reference context can also repel reproduced graphical values[28], while identical data can support opposing interpretations depending on which patterns capture attention[47]. Our findings further suggest that attention-directing choices not only influence how precisely individual values are judged, but whether they appear more extreme or more typical than they really are.

Similar concerns have been raised regarding political visualizations, where graphical context and attention can influence judgments about individual values. Furrer et al.[65] found that electoral maps using categorical red and blue colors led observers to overestimate the winning candidate's vote share, with smaller distortions when maps used continuous color gradients. Franconeri et al.[59] further emphasize that visualization design can direct attention toward particular values or comparisons through grouping, highlighting, and annotation, influencing which patterns observers extract from the same data. Our findings suggest that these attention-directing choices may also influence contextual bias. Focusing attention on an individual state could bias its estimated vote share away from the average of surrounding states, potentially exaggerating perceived differences. In contrast, viewers examining the map without focusing on particular states may rely more heavily on the overall distribution, making individual results less distinguishable from typical values. Therefore, how viewers direct their attention could influence whether individual states appear unusually different from, or more similar to the broader voting pattern.

Our findings also suggest testable ways to improve visualization design beyond choosing appropriate graphical encodings. For example, Franconeri et al.[59] recommend using highlighting and direct annotation to guide viewers toward relevant comparisons and reduce demands on working memory. Our results suggest such attention-directing features should also preserve access to exact values and their broader distribution, as emphasizing an individual item may itself introduce contextual bias. Interactive electoral maps could combine continuous color gradients with numerical labels showing a selected state's vote share and its position relative to other states. Similarly, searchable data points or linked tables could help viewers verify whether a particular value was present rather than relying on memory, which may favor values near the ensemble mean. In the overall context of previous findings, our results suggest that combining attention-guiding features with accessible numerical information may help distinguish genuine differences from contextual distortions, although these proposed safeguards require direct evaluation.

### C. Limitations and Open Questions

Several questions remain about the mechanisms and generalizability of our results. Most importantly, our no-cue condition combined the absence of target-specific information with a change in judgment from identifying a specified item to determining whether a value appeared anywhere in the set. Future work should disentangle attentional certainty from response framing to clarify whether these opposing biases arise during perception, memory, or decision-making[9].

In addition to uncertainty about which item is relevant, spatial and temporal uncertainty may also influence ensemble-induced bias. Temporal attention and expectation interact to affect visual discrimination[66], while rapid sequential presentation can limit memory for the order of individual items[67]. Such temporal constraints may also influence how information is integrated across successive items, particularly given evidence that ensemble representations are updated within discrete temporal windows[53]. Along these lines, Khayat et al.[32] found that attraction toward the ensemble mean was weaker when all items were presented simultaneously across different spatial locations than when they were presented sequentially. Future work independently manipulating featural, spatial, and temporal certainty is necessary to determine how each contributes to the magnitude and direction of ensemble-induced bias.

## VI. Conclusion

Previous research has shown that graphical context can improve the precision of individual-value judgments while introducing systematic biases[28]. Our findings extend this work by showing that the direction of ensemble-induced bias depends on attentional certainty. Across abstract shapes and real-world visualization contexts, specifying an individual target generally produced repulsion or negligible bias, whereas judgments without target-specific information produced attraction toward the ensemble mean and reduced precision. For visualization, attention-directing choices may therefore change not only which values viewers examine and how precisely they judge them, but also the direction in which surrounding data bias those judgments. When exact values matter, designs should provide direct access to them rather than relying on visual estimation or memory alone. The whole does more than represent its parts. It can warp them, and attention helps determine which way.

## ACKNOWLEDGMENT

Thanks to Ruth Rosenholtz, Jeremy Wolfe, and Arvind Satyanarayan for their comments on an early version of this work, and for Xingzhu (Brianna) Dong and Zihan (Emily) Fang for all their help with participant recruitment and data collection.